\documentclass[12pt,preprint,longabstract]{aastex}

\usepackage{amssymb}
\usepackage{amsmath}
\usepackage{epsfig}
\usepackage{graphicx}

\shorttitle{Magnetic depletions in the heliosheath}
\shortauthors{Drake et al.}

\begin{document}


\title{The formation of magnetic depletions and flux annihilation due to reconnection in the heliosheath}


\author{J.~F.~Drake\altaffilmark{1,2}, M.~Swisdak\altaffilmark{2}, M.~Opher\altaffilmark{3}, and J.~D.~Richardson\altaffilmark{4}}


\altaffiltext{1}{Department of Physics, the Institute for Physical Science and Technology and the Joint Space Institute, University of Maryland, College Park, MD 20742; drake@umd.edu}

\altaffiltext{2}{Institute for Research in Electronics and Applied Physics,, University of Maryland, College Park, MD 20742}

\altaffiltext{3}{Astronomy Department, Boston University, MA 02215}
\altaffiltext{4}{Kavli Center for Astrophysics and Space Science, Massachusetts Institute of Technology, Cambridge, MA 02139}


\begin{abstract}
The misalignment of the solar rotation axis and the magnetic axis of
the Sun produces a periodic reversal of the Parker spiral magnetic
field and the sectored solar wind. The compression of the sectors is
expected to lead to reconnection in the heliosheath (HS). We present
particle-in-cell simulations of the sectored HS that reflect the
plasma environment along the Voyager 1 and 2 trajectories,
specifically including unequal positive and negative azimuthal
magnetic flux as seen in the Voyager data \citep{Burlaga03}.
Reconnection proceeds on individual current sheets until islands on
adjacent current layers merge. At late time bands of the dominant flux
survive, separated by bands of deep magnetic field depletion. The
ambient plasma pressure supports the strong magnetic pressure
variation so that pressure is anti-correlated with magnetic field
strength. There is little variation in the magnetic field direction
across the boundaries of the magnetic depressions. At irregular
intervals within the magnetic depressions are long-lived pairs of
magnetic islands where the magnetic field direction reverses so that
spacecraft data would reveal sharp magnetic field depressions with
only occasional crossings with jumps in magnetic field direction. This
is typical of the magnetic field data from the Voyager spacecraft
\citep{Burlaga11,Burlaga16}.  Voyager 2 data reveals that fluctuations
in the density and magnetic field strength are anti-correlated in the
sector zone as expected from reconnection but not in unipolar regions.
The consequence of the annihilation of subdominant flux is a sharp
reduction in the ``number of sectors'' and a loss in magnetic flux as
documented from the Voyager 1 magnetic field and flow data
\citep{Richardson13}.

\end{abstract}





\section{INTRODUCTION}
\label{intro}
The rotation of the Sun twists the solar dipole field into the
dominant azimuthal Parker spiral magnetic field $B_T$ separated by the
heliospheric current sheet. Because of the misalignment of the solar
rotation axis and magnetic axis the current sheet flaps in the
vertical direction as it propagates outward from the Sun, producing
the sectored magnetic field \citep{Wilcox65} in which azimuthal
magnetic field $B_T$ reverses sign around every $13$ days (the RTN
coordinate system is defined with R in the radial direction, T in the
azimuthal direction with positive T in the direction of the Sun's
rotation and N, which points North in the equatorial plane, completes
the triad). The “sector-zone” occupies a latitudinal extent that
varies during the solar cycle, reaching nearly the poles when the
fields from the Sun are a maximum \citep{Smith01}.

An important question is whether the sectored magnetic field can
reconnect to release magnetic energy and accelerate particles. In the
solar wind around $1AU$ the heliospheric current only occasionally
undergoes reconnection \citep{Gosling07}, probably because the current
sheet is far wider than the characteristic ion inertial scale
$d_i=c/\omega_{pi}$ (where collisionless reconnection onsets)
\citep{Cassak05}. As a result, the sector stucture of the solar
magnetic field survives out to the termination shock (TS) even though
the periodicity of the current sheet becomes increasingly irregular
with distance from the Sun \citep{Burlaga03,Burlaga05,Burlaga08}. It
has been suggested that the drop in the solar wind density with
distance and therefore the increase in $d_i$, combined with the
compression of the current sheets downstream of the TS, leads to the
onset of reconnection in the sectored heliosheath (HS).  Such
reconnection has been proposed as a source of free energy to drive the
production of anomalous cosmic rays (ACRs) \citep{Drake10,Opher11} and
as a source of turbulence in the heliosheath that might control the
transport of energetic particles \citep{Burgess16}. The reduction of
the plasma flow in the HS on its approach to the heliospause is
expected to further compress the sectored magnetic field, inevitably
leading to the onset of reconnection in the HS
\citep{Czechowski10,Drake10,Opher11,Borovikov11}. However, determining the
thickness of the heliospheric current sheets downstream of the TS to
confirm that collisionless reconnection should onset in the HS is a
challenge because of the low and variable speed of plasma flows in the
HS and because weak magnetic fields there are difficult to measure. It
has been suggested that current sheets in the HS are thicker than the
ion inertial scale \citep{Burlaga11} but post reconnection final
states are characterized by magnetic islands in which current layers
are comparable in width to the island width. Thick current sheets
might therefore suggest that reconnection in HS has already taken
place.

In any case, smoking gun evidence from the Voyager observations that
reconnection in the sectored HS has taken place has not been
identified. The challenge is that the Voyager magnetometers were not
designed to measure the weak magnetic fields in the outer heliosphere
($\sim 0.1$nT with noise levels $\sim 0.05$nT) and Voyager 1 has no
plasma measurements. Further, the dynamics of reconnection in the high
$\beta$ environment of the HS remains relatively poorly understood
\citep{Schoeffler11,Schoeffler13} compared with the typically
$\beta\sim 1$ conditions at $1AU$. The multiple sharp dropouts of
energetic particles of heliospheric origin measured by Voyager 1 at
the HP boundary \citep{Stone13} do suggest the existance of magnetic
islands and therefore magnetic reconnection at and in the vicinity of
the HP\citep{Swisdak13,Strumik13,Strumik14}.

There are a variety of indirect indicators that reconnection is active
in the HS, including the loss in magnetic flux documented by Voyager 1
\citep{Richardson13} and the dropouts in the ``low energy'' electrons
at Voyager 2 \citep{Opher11,Hill14}. In the absence of reconnection
the azimuthal magnetic flux $V_RRB_T$ is preserved in the heliosphere
in regions where the flow is dominantly radial. While Voyager 1 was in
the HS, the radial flow $V_R$ dropped essentially to zero
\citep{Krimigis11} while there was no significant increase in $B_T$. A
significant flow in the N direction (to the North in the
case of Voyager 1) might convect the flux away and therefore prevent
the pileup of $B_T$. However, the Voyager 1 data suggested that $V_N$
was also very small \citep{Decker12,Stone12}. Thus, it seems likely that
reconnection must be playing a role in the Voyager 1 flux loss
measurements. On the other hand, reconnection actually does not
normally reduce the magnetic flux in a reconnecting current
layer. Rather, the flux that reconnects at an x-line is convected into
the adjacent magnetic island such that the integrated magnetic flux is
preserved \citep{Fermo10}. The flux loss issue can therefore not be
simply be resolved by invoking reconnection without a more careful
analysis. The dropouts in low energy electrons at Voyager 2 were
attributed to Voyager 2 leaving the sector zone
\citep{Opher11,Hill14}. The argument was that electrons can rapidly
escape from a region of the heliosheath with laminar (unreconnected)
magnetic fields while reconnected magnetic fields and associated
magnetic islands would be more effective in suppressing electron
transport along the ambient magnetic field. Thus, higher electron
fluxes in the sectored heliosheath are evidence that the heliosheath
magnetic field had reconnected -- in the absence of reconnection there
is no reason that the transport properties of the sectored and
non-sectored HS should differ.

The Voyager observations in the HS have uncovered other issues related
to the reconnection or not of the sectored magnetic field. The first
concerns the polarity of the HS magnetic field. The nominal polarity
of the magnetic field in the Northern hemisphere during the time period 2000-2011 is negative
(corresponding to the azimuthal angle $\lambda=\arctan
(B_T/B_R)=90^\circ$) \citep{Burlaga12,Richardson16}. When Voyager 1 is in the sector zone, $\lambda$
typically flips back and forth between $90^\circ$ and $270^\circ$ and
a long period of $90^\circ$, which occurred in 2011, would normally be
interpreted as an excursion out of the sector zone and into the
unipolar zone. On the other hand, the MHD models of the global
heliosphere predict that the sector zone should be convected to the
North as the HS plasma approaches the HP so that Voyager 1 should
remain within that sector zone prior to its crossing of the HP
\citep{Opher11,Borovikov11}. Similarly, in 2011-2012 Voyager 2 saw
significantly fewer excursions into negative (North latitude polarity)
than expected based on the Wilcox Solar Observatory predictions
\citep{Richardson16}. Why the Voyagers are seeing fewer excursions
into magnetic field polarities that are opposite to their
heliolatitude is a mystery.

A second mystery concerns the distinct magnetic structures seen in the
Voyager 1 and 2 data. At ``proton boundary layers (PBLs)'' the
magnetic field strength either rises or drops by factors of up to
three with no measureable change in the azimuthal angle $\lambda$ or
the elevation angle $\delta=\arcsin (B_N/B)$ (the
angle of the magnetic field with respect to the R-T plane)
\citep{Burlaga11,Burlaga12}. Since these regions of magnetic field
enhancement or depletion can last for days, it seems unlikely that
they are associated with kinetic scale instabilities such as mirror
modes, which typically produce localized ``humps'' in the magnetic
field rather than depletions in high $\beta$ plasma
\citep{Baumgartel03}.

Here we address the dynamics of reconnection in the sectored HS with
the goal of understanding the signatures and consequences of the
reconnection and the resultant structure of the HS magnetic field. The
simulations extend earlier models by considering more realistic
initial conditions that account for unequal positive and negative
azimuthal magnetic flux. That the sector zone does not carry equal
postive and negative flux at the latitudes of the Voyager spacecraft
trajectories is evident from the Voyager observations in the solar
wind in the outer heliosphere but upstream of the termination shock
\citep{Burlaga03}. In this region because the solar wind velocity
greatly exceeds the spacecraft velocity, time in a region of given
polarity is linked to the integrated magnetic flux in a given
sector. The Voyager data in the high-speed solar wind of the outer
heliosphere reveals that sector spacing is highly erratic and
therefore positive and negative fluxes are unequal. For the time
period 2000-2011 one might expect that for Voyager 1 the negative
polarity dominates because the spacecraft is closer to the Northern
boundary of the sector zone, which has negative polarity, while for
Voyager 2 positive polarity dominates. Unequal magnetic fluxes were required to reproduce Voyager 1 magnetic data in  MHD simulations of reconnection at the HP \citep{Strumik14}.

The consequence of unequal fluxes is profound. At late time when
reconnection is nearly complete, bands of single polarity flux
survive, which tends to organize the sector structure more than in
earlier simulations in which magnetic islands dominated the magnetic
structure at late time \citep{Drake10,Opher11}. The simulations are
also carried out with high initial $\beta$ and with initial force-free
current layers rather than Harris type current layers, which are
typically not seen even at 1AU \citep{Smith01}. Because of the high
$\beta$ the magnetic field at late time exhibits large-scale
depletions in which the magnetic field strength drops by around a
factor of three over a narrow boundary layer with little variation in
the direction of the magnetic field. The plasma density and
temperature rise slightly within the depletions to maintain pressure
balance. The radial width of these magnetic depletions is around three
times the width of the initial sector with the subdominant magnetic
flux. The depth of these depletions and their widths (when normalized
to the initial separation of current layers) are universal values that
are linked to the intrinsic properties of collisionless magnetic
reconnection. The boundaries of these magnetic depletions exhibit a
striking resemblance to the ``proton boundary layers'' seen in the
Voyager data. We show that reconnection on adjacent current layers
conspires to completely annihilate the subdominant magnetic flux,
leading to regions of unipolar flux. Pairs of magnetic islands do
survive at late time although the volume of plasma associated with
these remnant islands is small compared with the regions of magnetic
depletion. The island structures exhibit magnetic dips and rotations
in the magnetic field direction that might be interpreted as sector
crossings in satellite data. The simulations therefore offer a
possible explanation of the predominance of unipolar flux and the loss
of magnetic flux seen in the Voyager 1 magnetic field data. Finally,
analysis of the Voyager 2 magnetic field and plasma data reveals that
fluctuations in magnetic field strength and density are
anti-correlated in the sectored HS, as expected from reconnection, but
not in the unipolar HS. As a whole, the consistency of the Voyager
data with unique reconnection signatures establishes with high
likelihood that reconnection in the sectored HS has taken place.

\section{PIC model and initial conditions}
\label{picmodel}

We carry out 2-D particle-in-cell (PIC) simulations of the sector
structure in the $x-y$ plane of the simulation, which maps to the
heliospheric $R-T$ plane. The simulations are performed with the PIC
code p3d \citep{Zeiler02} using a periodic equilibrium magnetic field
\begin{equation}
  \begin{split}
  \frac{B_y}{B_0}=\tanh\left(\frac{x-0.2L_x}{w_0}\right)
  -\tanh\left(\frac{x-0.3L_x}{w_0}\right)\\
  +\tanh\left(\frac{x-0.7L_x}{w_0}\right)
  -\tanh\left(\frac{x-0.8L_x}{w_0}\right)-1.
  \end{split}
  \label{Bx0}
\end{equation}
For this magnetic configuration there are four current layers centered
at $x/d_i=20.48$, $30.72$, $70.68$ and $81.92$ on a computational
domain that is $L_y\times L_x=409.6\times 102.4$ where lengths are
expressed in units of $d_i=c_A/\Omega_i$ the proton inertial
length. The total magnetic flux in the positive $y$ direction is four
times that in the negative $y$ direction. The initial plasma density
$n_0$ and temperatures $T_e$ and $T_i$ are constants and the
out-of-plane magnetic field $B_z^2=B_0^2-B_y^2$ is chosen to produce
force balance.  This initial state is not a rigorous kinetic
equilibrium, especially for ions, but does not display unusual
behavior at early time. The results are presented in normalized units:
the magnetic field to the asymptotic value of the reversed field
$B_0$, the density to $n_0$, velocities to the proton Alfv\'en speed
$c_A=B_{0}/\sqrt{4\pi m_in_0}$, times to the inverse proton cyclotron
frequency, $\Omega_i^{-1}=m_ic/eB_{0y}$, and temperatures to
$m_ic_A^2$. We define some additional scale lengths as follows:
$w_0=0.25d_i$ is the half-width of an individual current sheet and
$\Delta_x=\Delta_y=0.05d_i$ are the grid scales. To maximize the
separation between the macroscales $L_x$ and $L_y$ and the kinetic
scales, we choose a modest ion to electron mass ratio of $25$ and
velocity of light $c$ of $15c_A$. The particle temperatures are
initially uniform with $T_i=5.0m_ic_A^2$ and $T_e=2.0m_ic_A^2$ so
$\beta=8\pi n_0(T_e+T_i)/B_0^2=14$ is large, as expected for the
heliosheath. The average number of particles per cell is
$100$. Reconnection begins from particle noise.

The overall scale sizes of our simulations are much smaller than those
of the heliospheric sectored field. The widths of the sectors upstream
of the TS are around $1.7\times 10^8km$, which at a density of
$0.001/cm^3$, is around $2\times 10^4d_i$.  Compression across the
shock and the approach to the heliopause reduces the sector width
somewhat but the sector spacing continues to be far larger in units of
$d_i$ than the values we can implement in our simulations. However, we
have shown earlier that the rate of growth of islands is insensitive
to the kinetic scale $d_i$ \citep{Schoeffler12} and the same
conclusion applies to the simulations presented here. Thus, the
reconnection rates and associated bulk ion flows can be translated to
the heliosheath by normalizing to the Alfv\'en speed and the Alfv\'en
transit time $L_x/c_A$.

\section{Simulation results}
\label{results}

\begin{figure}
  \epsscale{.5}
 \includegraphics[keepaspectratio,width=3.5in]{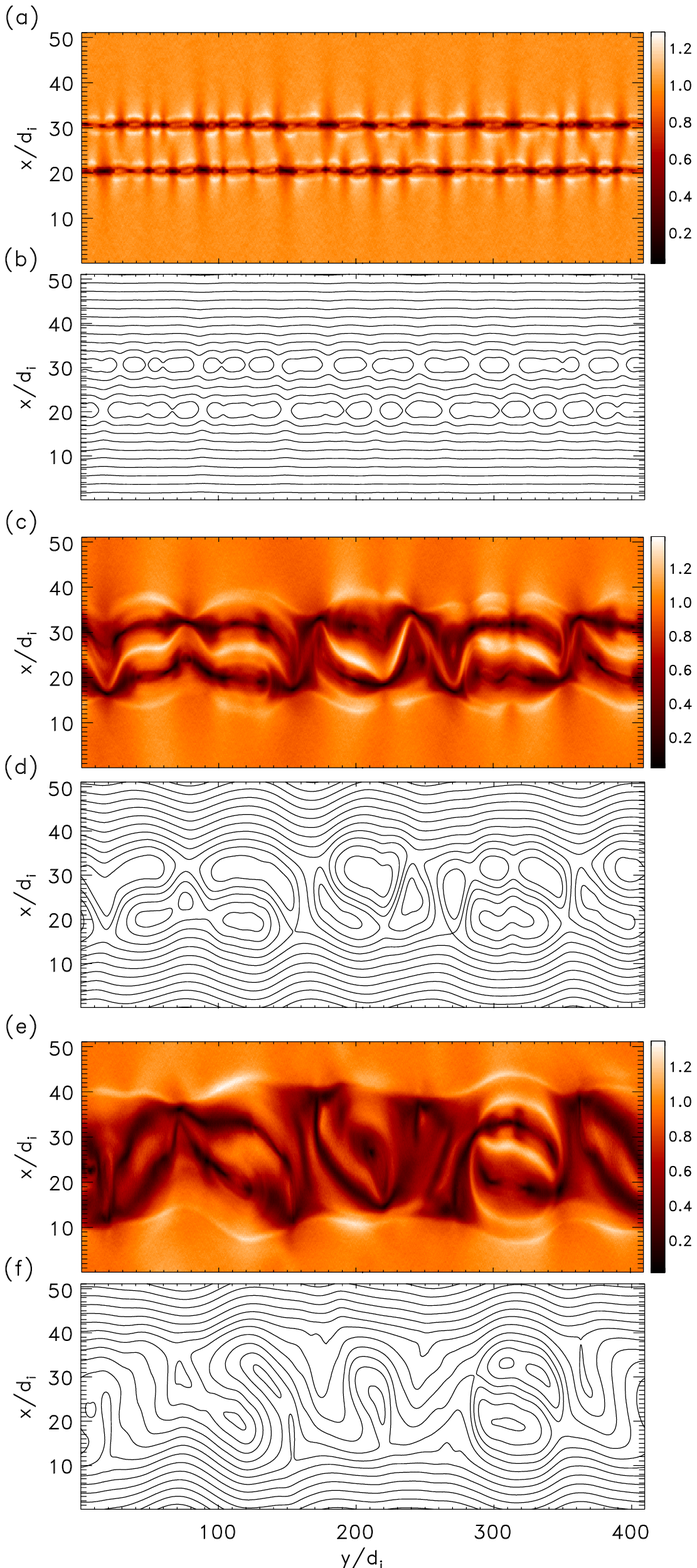} 
\caption{\label{psiB3t} (Color online) The magnetic structure in the
  lower half of the simulation domain at $\Omega_it=50$, $200$, and
  $350$ during the simulation. The magnetic field $B$ is shown in (a),
  (c) and (e) and the in-plane field lines are shown in (b), (d) and
  (f). The magnetic field $B$ from the simulation is available online as a video. The video runs from time zero to $\Omega_it=440.$ and shows the actual aspect ratio of the lower half of the computational domain.}
\end{figure}

In Fig.~\ref{psiB3t} we show 2D plots of the magnetic field strength
$B$ and the magnetic field lines around the two lower current layers
in the system at times $\Omega_it=50$, $200$ and $350$. The initial magnetic field points to the left above and below the two current layers and to the right between the two layers. In (a) and (b)
a large number of very small islands grow on the two current layers at
early time. These islands coalesce and continue to grow until in (c)
and (d) they span the entire region between the adjacent current
layers. At this point all of the magnetic field lines spanning the
simulation domain in the positive $y$ direction have reconnected. On
the other hand, positive magnetic flux still exists. Along a single
cut in $x$ in (d) $B_y$ has positive and negative values. However,
island merging and reconnection continues and at later time in (e) and
(f) much of the positive magnetic flux has been annihilated. How this
happens is important first because of the flux loss documented by the
Voyager 1 observations and second because magnetic reconnection
normally preserves the magnetic flux. At the magnetic x-line field
lines reconnect but the integrated unsigned flux through a magnetic
island is unchanged as reconnection proceeds. Namely, there is no flux
loss at the center of the island and magnetic flux is preserved
elsewhere so the total magnetic flux contained in an island is
preserved. On the other hand, it is evident from Fig.~\ref{psiB3t}(f)
that on a cut along $x$ at $y\sim 240$ $B_x$ will only have
positive values -- there is no surviving negative flux in this region.

The pressure anisotropy that develops during magnetic reconnection in
a high $\beta$ system such as the HS weakens the tension force exerted
by the magnetic field on the plasma and therefore slows reconnection
and allows islands take elongated forms
\citep{Drake10,Opher11,Schoeffler11}. A video animation of the
magnetic field $B$ for the full duration of the simulation domain (up
until $\Omega_{ci}t=440.$) is available online. The movie reveals that
the evolution of the magnetic field slows dramatically at late time, a
consequence of the pressure anisotropy, and allows isolated elongated
magnetic islands to survive late in time.

\begin{figure}
  \epsscale{.4}
  \includegraphics[keepaspectratio,width=2.7in]{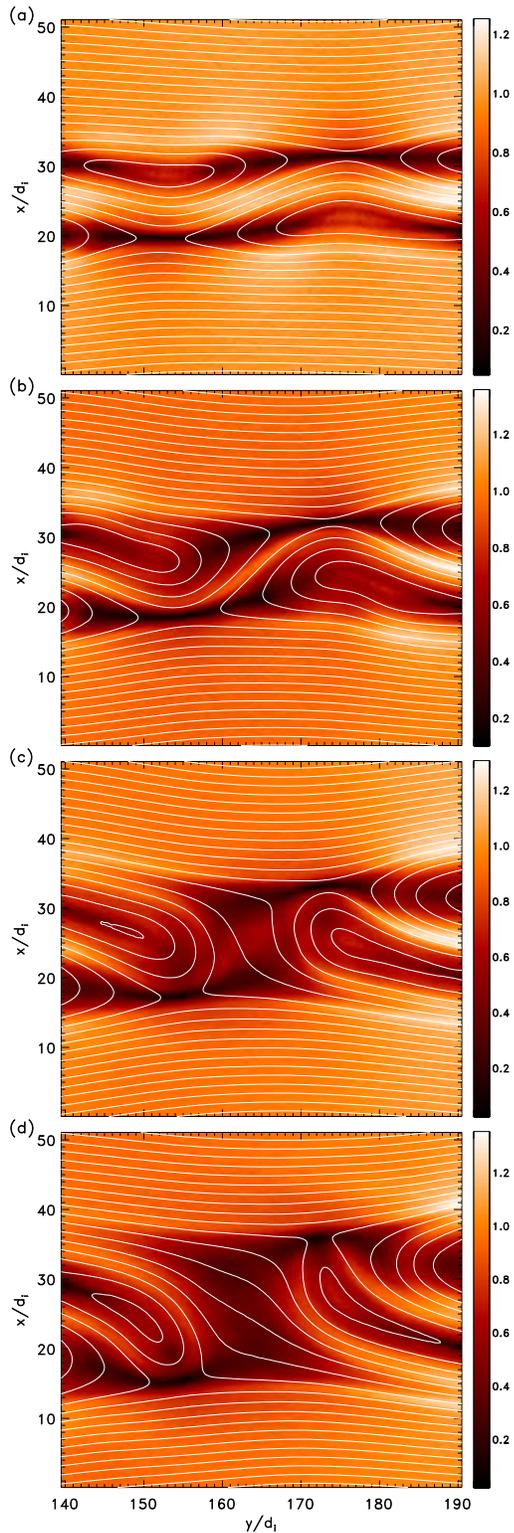}
\caption{\label{psiB4t} (Color online) A blowup view of the time evolution
  of a pair of magnetic islands at $\Omega_it=100$, $150$, $200$ and
  $250$ showing how the subdominant flux is
  annihilated. The magnetic field $B$ is shown in color and the overlaid
  white lines are magnetic field lines. The times shown are before the
  islands overlap the adjacent current layer in (a), when they first
  intersect the adjacent current layer in (b) and when the magnetic
  flux in the islands is eroded in (c) and (d).}
\end{figure}

In Fig.~\ref{psiB4t} we show a blowup of $B$ and associated magnetic
field lines at four times to illustrate how flux anniliation in the
geometry of the sectored heliosphere takes place. Again, the magnetic field lines above and below the two current layers initially point to the left and between the two current layers point to the right. In (a) magnetic
islands on the two current layers do not yet cross-connect with the
adjacent current layer and there is positive magnetic flux that spans
the domain in $y$.  In (b) the magnetic separatrix of the large island on the
lower current layer connects with the upper current layer. At this
point there is still substantial negative magnetic flux even though
there are no positive $B_y$ field lines that span the system along the
$y$ direction. However, the positive magnetic field $B_y$ of the
island on the lower current layer begins reconnecting with the
negative flux above the upper current layer, which annihilates the
positive flux. In (d) almost all of the magnetic flux in the island
from the lower current layer has been annihilated, eliminating the
surviving positive flux.

The late-time ($\Omega_pt=440$) structure of the magnetic fields,
density, and temperature of the full simulation domain are shown in
Figs.~\ref{cuts} and \ref{cuts_island}. In Fig.~\ref{cuts}(a) and (b)
are $B$ and field lines of the entire $x-y$ computational
domain. Surviving at late time are two pairs of islands embedded in a
lower and upper band of depleted magnetic energy. Elsewhere the
islands have reconnected away, leaving only negative magnetic
flux. The structure of the magnetic depletions is futher illustrated
in cuts of $B$ in (c), $n$ in (d), $T_i$ in (e) and the azimuthal
angle $\lambda=\arctan(B_y/B_x)$ in (f), where $\lambda$ defines the
direction of the magnetic field in the $x-y$ plane with respect to the
$x$ direction. The location of the cut is marked by the white line in
Fig.~\ref{cuts}{a}. The cut in Fig.~\ref{cuts}{a} is chosen
specifically because it represents what a satellite might typically
see in a heliosheath where reconnection has already annihilated
significant magnetic flux. The strong depletions of the magnetic field
seen in (c) span the region around each pair of adjacent current
layers in the initial system and are a consequence of growth and
merger of magnetic islands on adjacent current layers. The boundaries
of the magnetic depletions mark the maximum spatial extent of
reconnection of adjacent current layers. Pressure balance is
maintained within the magnetic depletions by small increases in the
density and ion temperature. The increase in electron temperature (not
shown) is even smaller. In the cut shown here the azimuthal angle
$\lambda$ is nearly constant at $270^\circ$. In this region all of the
initial positive flux has been annihilated. The scale length of the
boundaries of the magnetic depletions are expected to scale with the
proton Larmor radius $\rho_i$ because the protons carry most of the
pressure and they are able to decouple from the magnetic field on
scale lengths of the order of several $\rho_i$ \citep{Drake09a}. For
the simulation of Fig.~\ref{cuts}, the proton Larmor radius is around
$2.2d_i$ so the scale lengths of the boundaries of the magnetic
depletions in Fig.~\ref{cuts}(c) are $3-5\rho_i$. We emphasize that
the scale lengths of these boundary layers are insensitive to the
initial width of the initial current sheets because the boundaries are
associated with the upstream edges of the reconnection exhaust, which
is controlled by local physics -- ions move from upstream into the
exhaust and are accelerated up to the Alfv\'en speed across a narrow
boundary layer \citep{Drake09a}.

\begin{figure}
  \epsscale{.8}
  \includegraphics[keepaspectratio,width=6.0in]{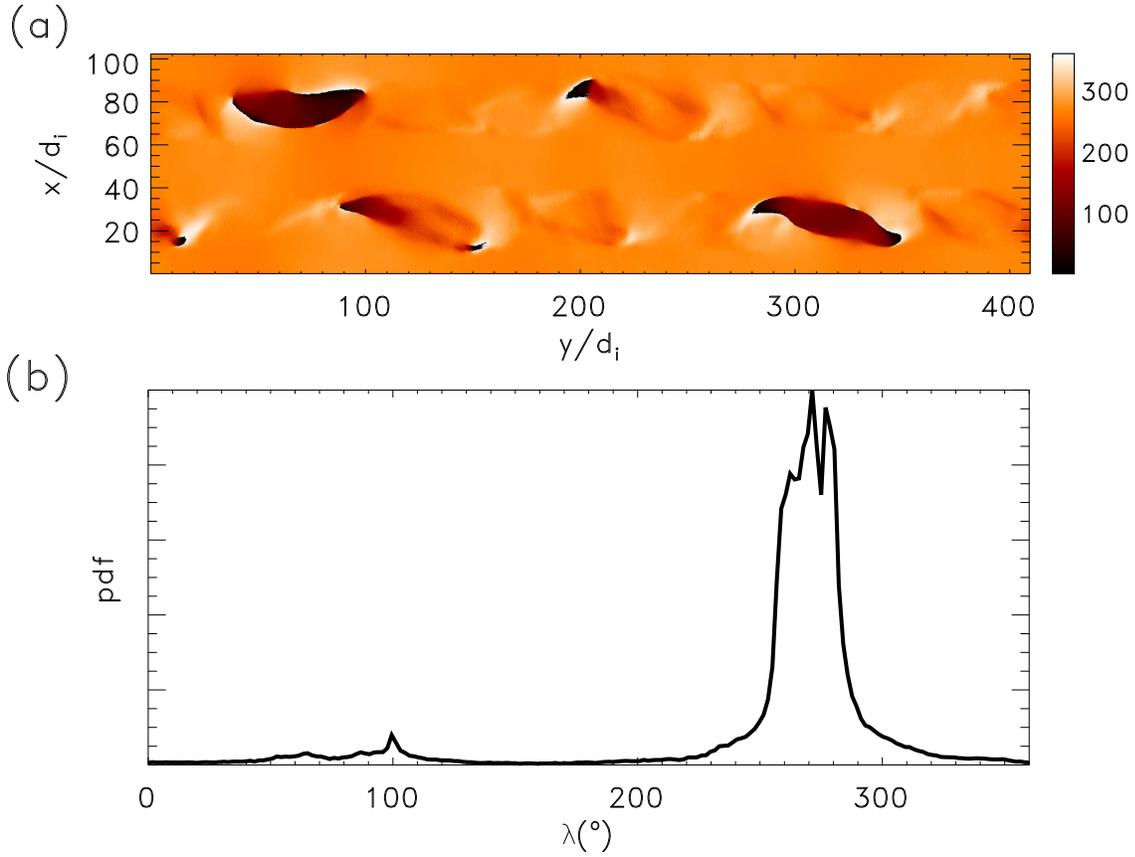}
\caption{\label{lambda} (Color online) At late time the spatial
  distribution of the azimuthal angle $\lambda$ in the full simulation
  domain (a) and its probability distribution in (b). Note the
  prominence of the dominant magnetic polarity at late time. The ratio
  probability of dominant to subdominant polarity in the initial state
  was four.}
\end{figure}

Figure \ref{cuts_island} is similar to Fig.~\ref{cuts} but the cuts
are now taken through the pair of magnetic islands on the top band of
depleted magnetic field. Again the white line shows the location of
the cut in Fig.~\ref{cuts_island}(a). The centers of the two islands
can be seen in Fig.~\ref{cuts_island}(b) but are most evident in the
cut of $\lambda$, which jumps sharply from $270^\circ$ to $90^\circ$ and back
across the centers of the islands. The island centers are locations of
minima of $B$ and small peaks in the density and ion
temperature. Distinguishing the crossing of such an island from the
crossing of the heliospheric current sheet in the heliosheath would be
difficult because the traditional signatures of reconnection, such as
high-speed flow, have died away -- the system has evolved to a
quasistatic state. We emphasize that the probability of crossing a
region where the subdominant magnetic flux survives at late time in
the system is small. In Fig.~\ref{lambda} the spatial distribution of
$\lambda$ at late time is shown in (a) and in (b) is the probability
distribution of $\lambda$. The probability of finding $\lambda\sim
90^\circ$ in this simulation at late time is finite but small.

\begin{figure}
  \epsscale{.8}
  \includegraphics[keepaspectratio,width=6.0in]{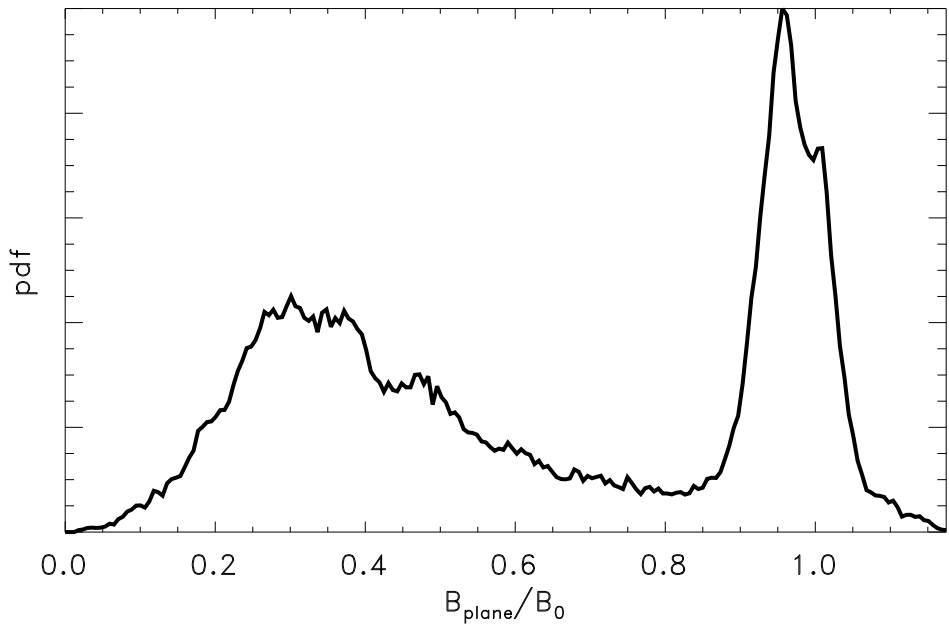}
\caption{\label{bperphist} From the simulation at late time the Probability Distribution Function (pdf) of the strength of the in-plane magnetic field $B_{plane}=\sqrt{B_x^2+B_y^2}$. Note the distinct peak at $B_0/3$.}
\end{figure}

We can give estimates for the widths and size of the magnetic
depletions based on how reconnection developed to produce the final
states shown in Figs.~\ref{cuts} and \ref{cuts_island}. The total
width of the magnetic depletion can be calculated by determining the
spatial extent of reconnection leading to the final state. Consider
two adjacent current layers separated by a distance $\delta L$ with a
total magnetic flux between the two layers $\delta\psi$. The
separatrix magnetic field line that connects to the two x-lines in
Fig.~\ref{psiB4t}(b) was originally at the center of the region
between the two nearby current layers at $x/d_i=20$ and $30$. Thus,
during the reconnection that led to this state half of the magnetic
flux in the region between the two current $\delta\psi$ layers
reconnected with the flux above the current layer at $30$ and half
reconnected with the flux below the current layer at $20$. Thus, the
spatial extent in $x$ of the entire region inside of the reconnected
field lines is $2\delta L$ ($\delta L/2$ above the upper current
layer, $\delta L/2$ below the lower current layer and the region
$\delta L$ between the two current layers). During the time from
Fig.~\ref{psiB4t}(b) to late time the remaining positive flux
$\delta\psi /2$ surrounding the island centered at $x\sim 20$ and
$y\sim 185$ reconnects with the negative flux above the upper current
layer. At the end of this reconnection process the island is gone. The
same happens as the positive flux in the island centered at $x\sim 30$
and $y\sim 145$ reconnects with the negative flux below the lower
current layer. The reconnection of these two islands with flux above
and below extends the reconnection zone another distance $\delta L/2$
above and below the reconnection zone shown in
Fig.~\ref{psiB4t}(b). Thus, at the end of the reconnection process the
total extension of the reconnection zone is $3\delta L$. The total
surviving magnetic flux over this domain is $\delta\psi$ ($\delta\psi$
each from above and below the initial current layers and $-\delta\psi$
from between the current layers). Since this flux is spread out over a
distance that is three times the initial separation of the current
layers, the magnetic field strength is $B_0/3$, independent of the
size of the simulation domain, current layer separation or plasma
parameters. The cuts of $B$ in Fig.~\ref{cuts}(c) display depletions
with widths that are around $30$, consistent with this estimate, and
with magnetic field minima that are close to $0.33B_0$. A Probability
Distribution Function (pdf) of the magnitude of the in-plane magnetic
field $B_{plane}$ is presented in Fig.~\ref{bperphist}. There are
distinct peaks in the pdf at the initial magnetic field strength $B_0$
and at $B_0/3$. Of course, if the disparity between positive and
negative flux is insufficient, the magnetic islands growing on all
current layers will ultimately overlap and the organized magnetic
depletions shown in Figs.~\ref{cuts} and \ref{cuts_island} will not
characterize the final state. The disparity between positive and
negative flux needs to exceed two for the isolated depletions to
survive at late time.

\section{Observational Results from Voyager 2}

Our simulations of the sectored HS suggest that large depletions in
the magnetic field result from magnetic reconnection. The ``proton
boundary layers'' \citep{Burlaga11,Burlaga12} that have been
identified in the Voyager 1 and 2 data look very much like the
boundaries of the strong magnetic depletions that we see in our
simulations. Because reconnection-driven flows are Alfv\'enic and
therefore fall below the local magnetosonic velocity in the
high-$\beta$ heliosheath, reconnection dynamics is to lowest order
incompressible, which means that depletions in the magnetic field
pressure correspond to enhancements in the plasma density and
pressure. This can be seen in the cuts across the magnetic depletions
in the simulation data presented in Figs.~\ref{cuts} and
\ref{cuts_island}. Thus, if the magnetic field disturbances in the
sectored heliosheath are driven by reconnection, they should
correspond to perturbations in the density such that deviations of the
magnetic field and density from the ambient background are
anti-correlated -- reductions (increases) of the magnetic strength
should correspond to local increases (decreases) in the plasma
density. In the non-sectored heliosheath this anti-correlation should
not be present unless there are mechanisms for generating
non-compressible turbulence other than reconnection.

\begin{figure}
  \epsscale{.8}
  \includegraphics[keepaspectratio,width=4.0in]{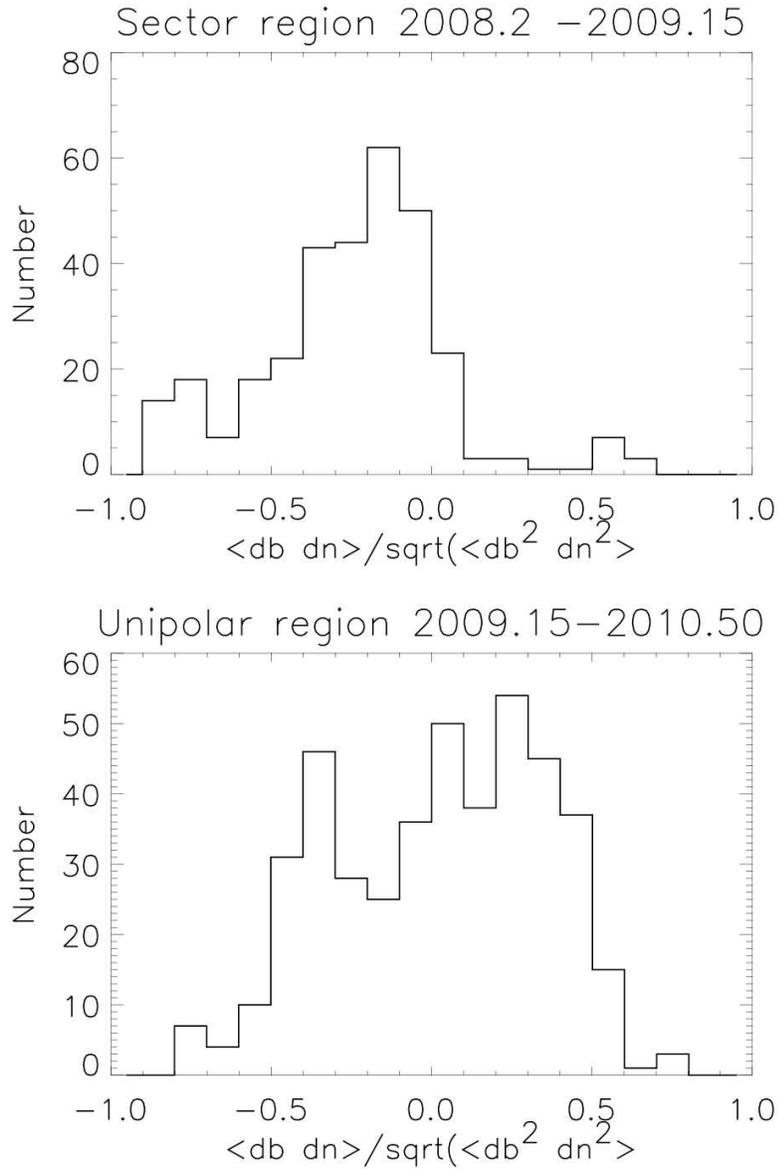}
\caption{\label{hist} Histograms of $<dBdn>/\sqrt{<dB^2><dn^2>}$ where $dB
= B - <B>$ and $dn = n - <n>$ from Voyager 2 in the sectored region (top) and the unipolar region (bottom). We use 25 day averaging windows.}
\end{figure}

Thus, we have explored the correlation between magnetic field and
density fluctuations in the Voyager 2 datasets. The plasma instrument
on Voyager 1 failed many years ago so correlation studies with the
Voyager 1 data are not possible.  We have compared the fluctuations of
the density ($dn$) and magnetic field magnitude ($dB$) in a
heliosheath region where the sector structure is observed (2008.2 -
2009.15, 344 days of data) with those in a unipolar region (2009.15
-2010.5, 455 days of data) \citep{Burlaga11,Richardson16}. We use
daily averages of the magnetic field magnitudes from the SPDF website
and of the density from the MIT Voyager web site. Figure \ref{hist}
shows histograms of $<dBdn>/\sqrt{<dB^2><dn^2>}$ where $dB = B - <B>$
and $dn = n - <n>$. The average $<A>$ of any quantity $A$ is defined over a $25$ day averaging window. The histograms
are clearly different for the two regions, with $dB$ and $dn$ usually
having opposite signs (anti-correlated) in the sector region but not
in the unipolar region. Thus, the Voyager 2 data suggests that
fluctuations in the sectored heliosheath are driven by
reconnection. Alternative explanations for the anti-correlation
between the fluctuations in the magnetic field intensity and density
would need to explain why the fluctuations are anti-correlated in the
sectored zone but not in the unipolar region.

\begin{figure}
  \epsscale{.8}
  \includegraphics[keepaspectratio,width=6.0in]{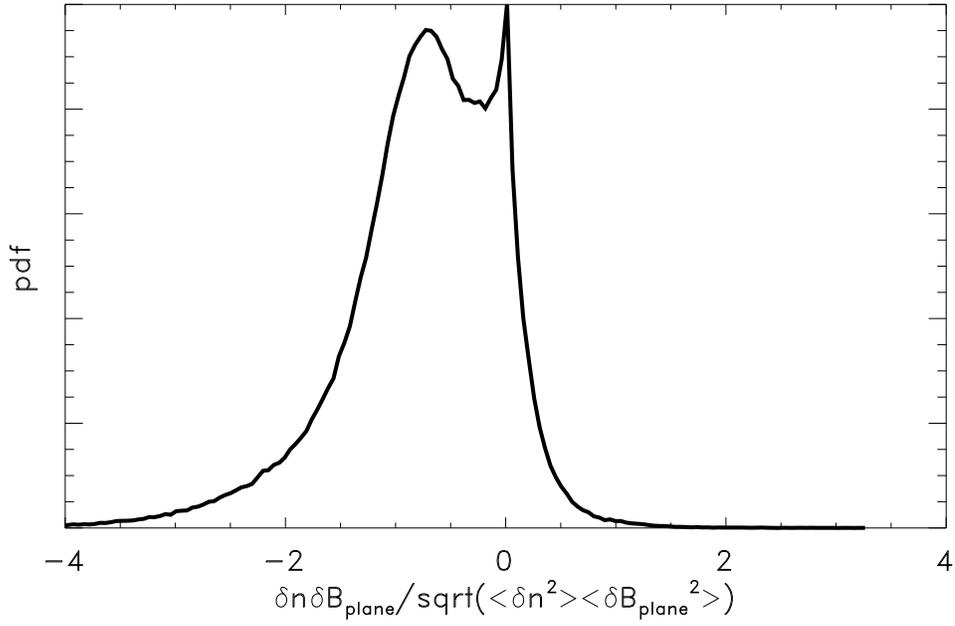}
\caption{\label{bnhist} From the simulation a histogram of
  $\delta n\delta B_{plane}/\sqrt{<\delta B_{plane}^2><\delta n^2>}$ where $\delta n = n - <n>$, $\delta B_{plane} = B_{plane} - <B_{plane}>$ with $n$ the density, $B_{plane}$ the magnitude of the in-plane magnetic field and $<n>$ and $<B_{plane}>$ the average over the computational domain. }
\end{figure}

To compare this correlation data with that from our simulation, we
evaluate
\begin{equation*}
  \frac{\delta n\delta B_{plane}}{\sqrt{<\delta n^2><\delta B_{plane}^2>}}
 \end{equation*}
with $\delta f=f-<f>$ for any function $f$ and where
$<f>$ is an average over the simulation domain. The pdf of this
correlation is shown in Fig.~\ref{bnhist}. As in the observational
data the density and $B_{plane}$ are anti-correlated and there is a
distinct tail on the negative side of the distribution.
\section{Discussion}
\begin{figure}
  \epsscale{.5}
  \includegraphics[keepaspectratio,width=3.5in]{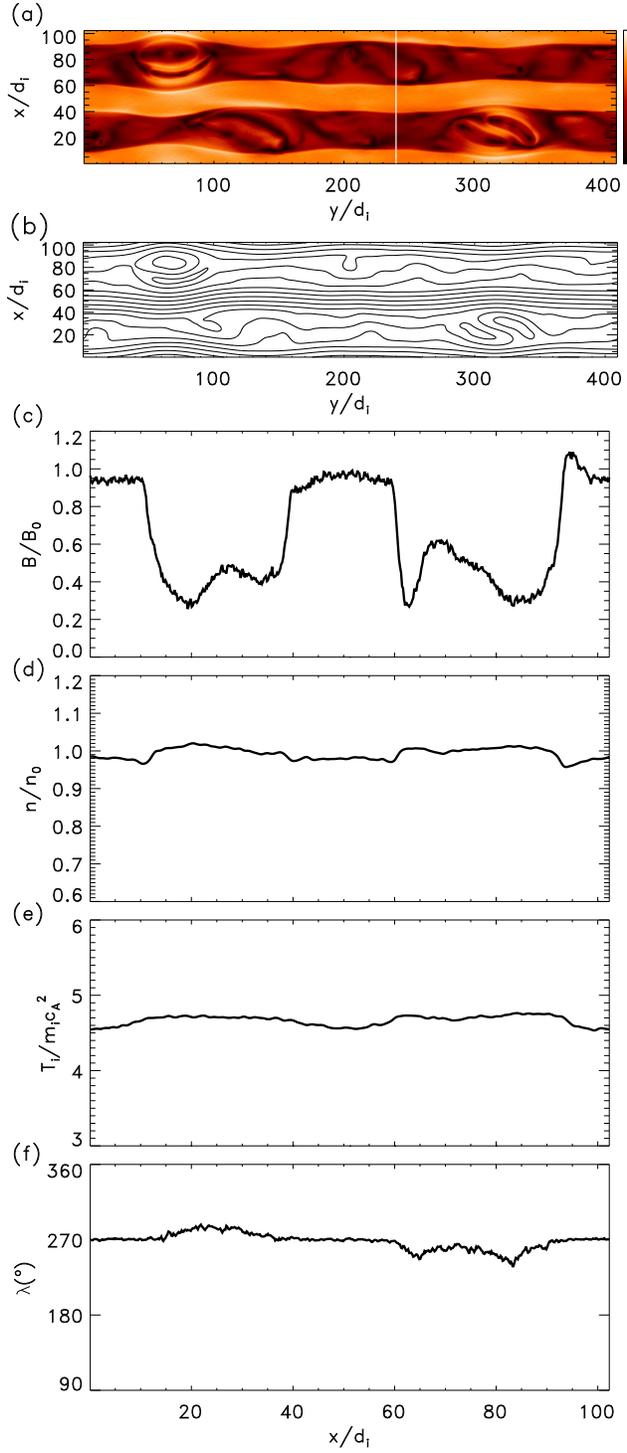}
\caption{\label{cuts} (Color online) The late-time
  ($\Omega_it=440$) structure of the magnetic field $B$ in (a) and
  magnetic field lines in (b) over the entire simulation domain. Cuts
  across the magnetic depletions along the white line in (a) showing
  $B$ in (c), of the density $n$ in (d), the ion temperature $T_i$ in
  (e) and the azimuthal angle $\lambda$ in (f). }
\end{figure}

\begin{figure}
  \epsscale{.5}
  \includegraphics[keepaspectratio,width=3.5in]{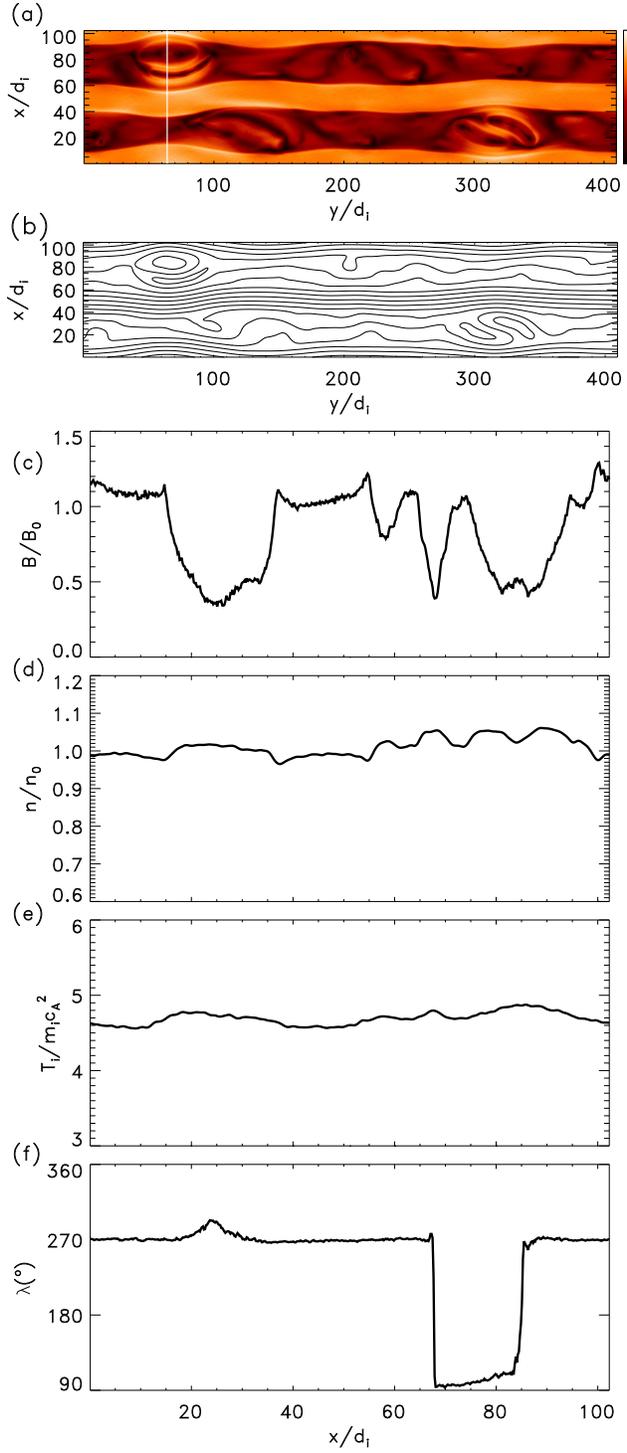}
\caption{\label{cuts_island} (Color online) The late-time
  ($\Omega_it=440$) structure of the magnetic field $B$ in (a) and
  magnetic field lines in (b) over the entire simulation domain. Cuts
  through a remnant pair of magnetic islands along the white line in (a) showing
  $B$ in (c), of the density $n$ in (d), the ion temperature $T_i$ in
  (e) and the azimuthal angle $\lambda$ in (f). }
\end{figure}

We have carried out kinetic simulations of magnetic reconnection in
the sectored heliosheath that include the asymmetry in the magnetic
flux in the sectors that is expected near the Northern and Southern
latitude boundaries of the heliospheric sector zone. We show that when
the magnetic flux asymmetry is more than a factor of two, bands of
unreconnected magnetic flux $B_T$ survive and sandwich bands that have
strongly depleted magnetic field. In the final state the widths of the
depletion regions are around three times the width of the initial
current layer separations with magnetic field strengths that are
around one third of the pre-reconnection intensity. The boundaries of
the depletion regions are sharp -- several times the proton Larmor
radius. The reconnection of magnetic islands on adjacent current
sheets leads to the nearly complete annihilation of flux in the
subdominant direction. Such flux annihilation does not typically take
place during magnetic island growth on a single current layer because
magnetic islands conserve the integrated magnetic flux. The final
state has scattered pairs of remnant magnetic islands in which the
subdominant magnetic flux survives. Cuts across these islands would
appear like crossings of an undisturbed heliospheric current sheet --
the azimuthal angle $\lambda$ jumps sharply from $90^\circ$ to
$270^\circ$ and then reverses across the cores of these islands
(Fig.~\ref{cuts_island}). The probability of crossing such an island,
however, is low compared to that of crossing a pristine region of
magnetic depletion (Fig.~\ref{cuts}).

In the spherically expanding solar wind the conservation of magnetic
flux implies that $V_RB_TR$ is a constant. For a constant solar
velocity $V_R$ this expression yields the usual falloff of $B_T$ as
$1/R$, which has been documented in the solar wind. In the heliosheath
the magnetic field is more complex because of the development of
latitudinal flows $V_N$. However, it was a major surprise when the
estimated radial plasma flows $V_R$ at the Voyager 1 spacecraft
dropped to essentially zero in 2010 \citep{Krimigis11} and yet the
magnetic field strength did not increase to compensate
\citep{Burlaga12}. Since the velocity $V_N$ was also close to zero,
the loss of magnetic flux through a flow to high latitude was
insufficient to explain the apparent loss of flux. The conclusion
therefore was that the Voyager 1 observations documented a flux loss
in the heliosheath \citep{Richardson13}. In contrast, the magnetic
field and flow measurements at Voyager 2 suggested that flux was
conserved along its trajectory.

The model presented here might explain how magnetic flux could be lost
in the sectored heliosheath. Of course, if flux annihilation as
discussed here did take place in the heliosheath at the location of
Voyager 1, one would expect to see fewer sector crossings than
expected in the magnetic field data. From 2010 to the heliopause crossing
in mid 2012 the Voyager 1 spacecraft did see less Southern polarity
flux than expected from the WSO data. However, during the long period
during which the measured value of $V_R$ decreased at Voyager 1, the
data does not suggest a reduced probability of Southern polarity
magnetic flux. Indeed, there is an unexplained period during 2008-2010
when the probability of seeing Southern polarity flux is much higher
than expected \citep{Richardson16}. Magnetic reconnection nevertheless
seems to be the only viable mechanism that can explain the flux loss
along the Voyager 1 trajectory.

The challenge is to identify a more direct method of establishing
whether reconnection is taking place in the heliosheath. This is not
easy because crossing a current sheet where reconnection is actually
taking place is highly improbable. There have now been several hundred
identifications of active reconnection in the solar wind at 1AU and
yet there is not a single documented observation of the crossing of a
magnetic x-line where active reconnection is ongoing -- how does one
distinguish a static current layer from a current layer where
reconnection is active? This requires the accurate measurement of the
intense Hall electric field that bounds the current layer on either
side of the x-line \citep{Drake08}, which has not been measured in the
solar wind. The documented reconnection observations
\citep{Gosling05,Gosling07} are crossings of the reconnection exhaust,
where the measured exhaust velocity has been cross-checked with the
predictions based on the Wal\'en condition \citep{Hudson70}. What are
the corresponding direct signatures of reconnection in the
heliosheath? We have argued previously that if reconnection onsets
just downstream of the termination shock, where the heliospheric
current sheet should be compressed below the ion inertial scale $d_i$,
the growth time for islands to reach the characterisic sector spacing
should be around 60 days, which translates to a distance around 2-3 AU
downstream of the termination shock \citep{Schoeffler12}. Deep within
the heliosheath the sectors should therefore have reconnected and the
signatures of reconnection should reflected in the late-time structure
of the magnetic remnants of reconnection rather than an active
reconnection site. That reconnection is taking place in the
heliosheath is also supported by the ACR spectra, which peak well
downstream of the termination shock
\citep{Stone05,Decker05,Decker08}. Reconnection downstream of the
termination shock is one possible explanation of these observations
\citep{Drake10,Opher11}. Some more recent theoretical models suggest
that reconnection dynamics downstream of the shock is an intrinsic
component of the termination shock structure and associated particle
acceleration \citep{Zank15}.

In the core of the sector zone, where the positive and negative
polarity fluxes are nearly equal, the late time state consists of
elongated magnetic islands \citep{Opher11} while closer to the
latitudinal boundaries of the sector zone where positive and negative
polarity fluxes are not equal, the spacecraft observations should
resemble the cuts shown in Figs.~\ref{cuts} and \ref{cuts_island}. The
magnetic depletions across rather narrow boundary layers, which can
extend to several AU in width, are the clearest direct signatures of
the post reconnection heliosheath. The ``proton boundary layers
(PBLs)'' that have been documented in the Voyager 1
\citep{Burlaga11,Burlaga12} and 2 \citep{Burlaga16} data are very
similar to the magnetic depletions that appear in our simulations. A
surprise is that many of the clearest examples of PBLs from the
Voyager 1 data seem to have jumps in magnetic field strength that are
around three, which is the value which follows from our analysis of
the reconnection dynamics. The measured scale lengths of the measured
PBLs are $5-10\rho_i$, which are modestly wider than those seen in our
simulations. Other suggestions are that the PBLs result from the
growth of mirror modes \citep{Burlaga11,Burlaga12}. However, mirror
modes tend to take the form of humps rather than depletions in high
$\beta$ systems \citep{Baumgartel03} and the overall spatial scale of
the depletions from mirror modes are far smaller than the typical
depletions measured in the heliosheath. In contrast the size of the
depletions from reconnection are not linked to any kinetic scale but
to the radial scale length of the sectors, which can be of the order
of an AU. The depletions from mirror modes would also require a
significant magnetic field component in the N direction so that the
elevation angle $\delta$ would be substantial. Large values of
$\delta$ that extend over the regions of magnetic field depletion are
not typically seen in the spacecraft data.

Finally, intrinsic to reconnection in a high $\beta$ system such as
the HS, where the Alfv\'en speed is well below the magnetosonic speed,
is that the dynamics is nearly incompressible so that the magnetic
depletions seen in the simulations are supported by corresponding
increases in the plasma pressure (and density). Thus, the fluctuations
in the magnetic field strength and density from reconnection should be
anti-correlated. The data from Voyager 2 confirms the anti-correlation
of fluctuations in magnetic field strength and density in the sectored
HS but not in the unipolar HS as expected if reconnection is taking
place in the sectored HS. 

\begin{acknowledgements}
This work has been supported by NASA Grand Challenge NNX14AIB0G, NASA
awards NNX14AF42G, NNX13AE04G and NNX13AE04G, and NASA contract 959203
from JPL to MIT.  The simulations were performed at the National
Energy Research Scientific Computing Center. We acknowledge fruitful
discussions with Dr.\ Len Burlaga on the Voyager observations and with
Dr.\ Obioma Ohia on outer heliosphere reconnection. This research
benefited greatly from discussions held at the meetings of the
Heliopause International Team “Facing the Most Pressing Challenges to
Our Understanding of the Heliosheath and its Outer Boundaries'' at the
International Space Science Institute in Bern, Switzerland.
\end{acknowledgements}


\clearpage

\end{document}